\documentstyle[psfig]{lamuphys}
\makeatletter
\let\chapter\hid@chapter
\def\etal{et~al.}
\def\spose#1{\hbox to 0pt{#1\hss}}
\def\lta{\mathrel{\spose{\lower 3pt\hbox{$\mathchar"218$}}
     \raise 2.0pt\hbox{$\mathchar"13C$}}}
\def\gta{\mathrel{\spose{\lower 3pt\hbox{$\mathchar"218$}}
     \raise 2.0pt\hbox{$\mathchar"13E$}}}
\def\Ha{H$\alpha$}
\makeatother
\newenvironment{bul}
{\begin{list}
  {$\quad \bullet$}
  {\itemsep = 0.3ex\parsep=0pt\topsep = 0mm}}
{\end{list} }
\begin{document}
\pagenumbering{arabic} 

\title{The 3CR radio galaxies at $z \sim 1$: \\ Old stellar populations in
central cluster galaxies}

\titlerunning{The 3CR radio galaxies at $z \sim 1$}

\author{Philip\,Best\inst{1,2}, Malcolm\,Longair\inst{2}, and Huub
R\"ottgering\inst{1}} 

\institute{Sterrewacht Leiden, Huygens Lab, Postbus 9513, 2300\,RA,
Leiden, The Netherlands
\and
Cavendish Laboratory, Madingley Road, Cambridge CB3 0HE, England}

\maketitle

\begin{abstract}

We investigate the old stellar populations of the 3CR radio galaxies at
redshift $z \sim 1$ using observations made with the Hubble Space
Telescope and the United Kingdom InfraRed Telescope. At radii $r \lta
35$\,kpc, the infrared radial intensity profiles of the galaxies follow de
Vaucouleurs' law, whilst at larger radii the galaxies show an excess of
emission similar to that of low redshift cD galaxies. The locus of the
high redshift 3CR galaxies on the Kormendy relation is investigated:
passive evolution of the stellar populations is required to account for
their offset from the relation defined by low redshift giant ellipticals
and brightest cluster galaxies. The 3CR galaxies, on average, possess
larger characteristic radii than the low redshift brightest cluster
galaxies. Coupled with existing evidence, these results are strongly
suggestive that distant 3CR galaxies must be highly evolved systems, even
at a redshift of one, and lie at the centre of moderate to rich
(proto--)clusters.
\end{abstract}

\section{Introduction}

The revised 3CR sample of radio sources defined by Laing~\etal\ (1983)
consists of the brightest radio sources in the northern sky, selected at
178\,MHz. It contains radio galaxies and quasars out to redshifts $z \sim
2$. The low redshift radio galaxies in the sample have long been known to
be associated with giant elliptical galaxies containing old stellar
populations; if the high redshift sources are similarly associated with
giant ellipticals then these sources provide an ideal opportunity to study
the evolution of stellar populations at early cosmic epoch, and thus to
constrain models of galaxy formation and evolution.

Following the advent of infrared bolometers, Lilly and Longair (1982,1984)
obtained infrared K--magnitudes for an almost complete sample of 83 3CR
galaxies with redshifts $0 < z < 1.6$.  Plotting K magnitude against
redshift for these objects, they showed that the resulting relation has a
remarkably small scatter ($\lta 0.6$ magnitudes). The tightness of this
correlation was interpreted as indicating that the high redshift 3CR host
galaxies are also giant elliptical galaxies.  Lilly and Longair (1984)
showed that, unless the deceleration parameter is as large as $q_0 \sim
3.5$, the shape of the K$-z$ relationship is not consistent with
non--evolving stellar populations, but that at least passive evolution is
required.

The optical--ultraviolet morphologies of the distant galaxies are,
however, far more complicated. In 1987, McCarthy \etal\ and Chambers
\etal\ discovered that the optical emission of powerful high redshift
radio galaxies tends to be elongated and aligned along the direction of
the radio emission. Our HST images have allowed us to study the
morphologies of these galaxies on kpc scales, and demonstrate that the
form of the alignment differs greatly from source to source (Longair
\etal\ 1995, Best \etal\ 1996,1997b); in some cases it arises from the
elongation of a single central emission region, whilst in others strings
of bright knots are seen to stretch along the radio axis. The most
promising models for this aligned emission are star formation induced by
the radio jet, scattering of quasar light by dust or electrons, or nebular
emission from the warm ionised gas (eg. R\"ottgering and Miley 1996).

\section{Spectral energy distributions}

The critical point about all of the mechanisms for producing the aligned
emission is that they possess relatively flat spectra, and so at the
longer wavelengths of the infrared emission they are dominated by the
emission of the underlying old stellar population.

To estimate the fraction of the K--band flux density that would arise from
the aligned emission, we produce a relatively simple model for the
spectral energy distributions (SED's) of the galaxies, using a combination
of two components: (i) a passively evolving old stellar population, and
(ii) a flat spectrum ($f_{\nu} \propto \nu^0$) emission component,
representing the aligned emission. The SED of the first component was
derived using the stellar synthesis codes of Bruzual and Charlot
(1993,1997), assuming that the stars formed in a 1\,Gyr burst beginning at
a redshift $z=10$. A Scalo IMF with upper and lower mass cut--offs of 0.1
and 65$M_{\odot}$ was adopted. The precise spectral shape of the second
component of the fit depends upon the nature of the alignment effect; the
adoption of a flat spectral index provides a good compromise and will not
be too far wrong in any case.

\begin{figure}[t]
\centerline{\psfig{figure=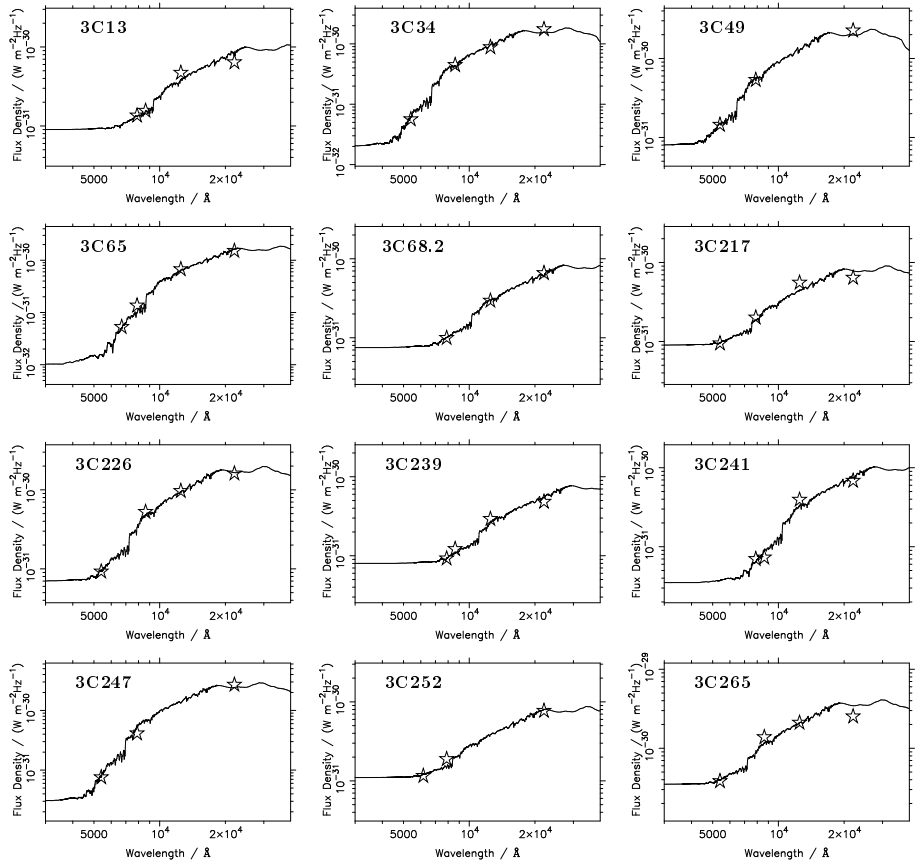,width=\textwidth,clip=}}
\caption{\label{specfit} SED fits to the broad band flux densities of the
3CR galaxies using an old stellar population and a flat spectrum component
(see text for details).}
\end{figure}

For each galaxy, we calculate the sum of these two components which best
matches the broad band flux densities of the galaxies measured at four
different wavelengths from our UKIRT and HST images, taking account of any
emission line contributions (Best \etal\ 1997a). 3C437 and 3C470, for
which only two broad band flux densities were available, and 3C22 and 3C41
which both possess strong nuclear components in the infrared images (see
Section~3) were omitted from this analysis. The fits obtained are
generally good. To illustrate this, in Figure~\ref{specfit} we show
the results for the first half of the sample. These indicate that the
simple two component model provides a good representation of the SED's of
these galaxies. The percentage of K--band light associated with the flat
spectrum component ranges from only $\sim 1\%$ in the very passive sources
3C65 and 3C337, up to $\sim 22\%$ in the case of 3C368, with an mean value
of $\sim 8\%$. Although these percentages would be higher if a redder
spectral shape had been adopted for the aligned component, the assumption
that the K--band light is dominated by the old stellar population seems
reasonably secure.

\section{Radial intensity profiles}

\begin{figure}[t]
\centerline{
\psfig{figure=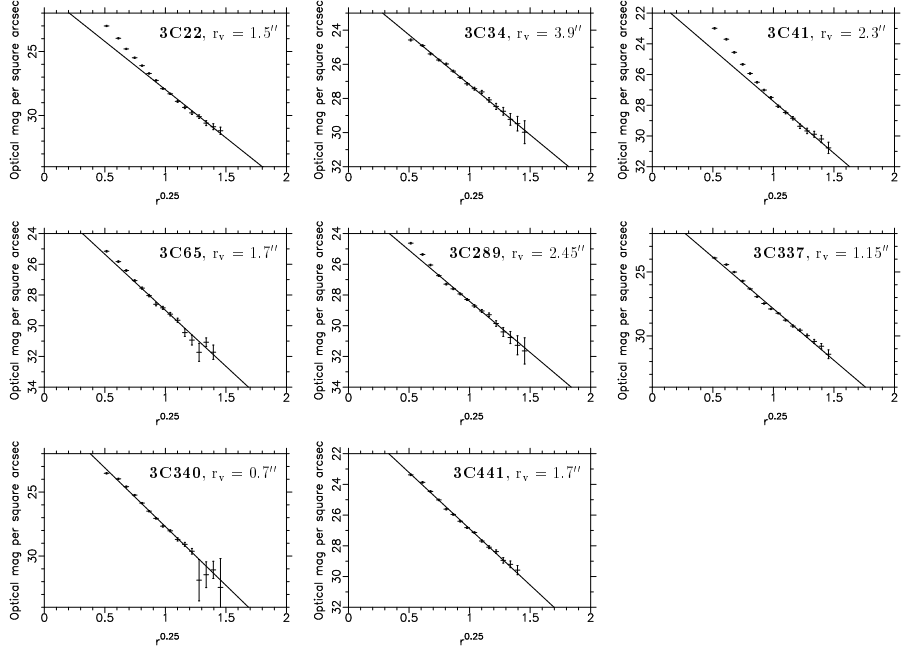,clip=,width=\textwidth,angle=-90}
} 
\caption{\label{hstvaucs} De Vaucouleurs fits to the radial intensity profiles
of the HST images of eight 3CR radio galaxies which do not show a
significant active ultraviolet component. The characteristic radius of
each, determined from the gradient of the best fitting straight line, is
given.}
\end{figure}

Our observations can be used to compare the radial intensity profiles of
these galaxies with de Vaucouleurs' law: $I(r) \propto {\rm exp} \left
[-7.67 \left (r/r_{\rm e}\right )^{-{1/4}}\right ]$. For 8 of the galaxies
in the sample there is little evidence for a significant ultraviolet
emission component, in the sense that only a small ($\le 5 \mu$Jy) flat
spectrum component is required in the fit to their SED, and the
morphologies of the HST images are almost symmetrical. For these galaxies,
the radial profiles of the optical emission were measured, with nearby
companions objects being removed and replaced by the average of the
background pixels at that distance from the centre of the galaxy. The
radial profiles are shown in Figure~\ref{hstvaucs}: in 6 of the 8 cases a
de Vaucouleurs profile provides an excellent match to the observed
data. The cases of 3C22 and 3C41 are discussed below. The values of the
characteristic radius can typically be measured to an accuracy of order
15\%.

\begin{figure}[t]
\centerline{
\psfig{figure=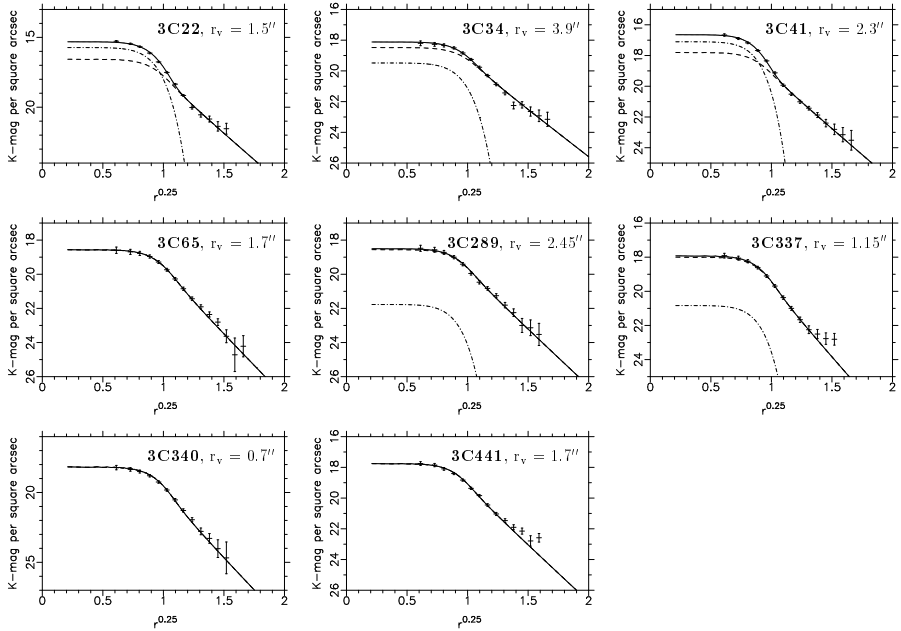,clip=,width=\textwidth,angle=-90}
}
\caption{\label{kvaucs} Fits to the radial intensity profiles of the
K--band images of the eight 3CR galaxies shown in Figure~\ref{hstvaucs},
using the sum of an unresolved point source (dash--dot line) and a de
Vaucouleurs profile with the characteristic radius determined from the HST
images (dashed line). For each of the profiles, the effect of seeing has
been taken into account. The sum of the two components is indicated by the
solid line. (Note that in many cases the best fit does not involve a point
source component.)}
\end{figure}

Using the characteristic radii determined from the HST images, the
infrared profiles of these galaxies can also be investigated. For each of
the galaxies, a de Vaucouleurs profile with the characteristic radius
derived from the HST fit, and an unresolved emission source were each
convolved with effects of the seeing (which was typically between 1 and
1.2 arcsec). The combination of these two components which provided the
best match to the observed data was then determined in each case, the
results being shown in Figure~\ref{kvaucs}: the dashed line shows the
radial profile of the de Vaucouleurs component, the dotted line shows that
of the point source component, and the solid line shows the total
intensity profile (note that in many cases, the best fit was produced
using no point source component, and so this does not appear on the
plot). In each case a good match is obtained.

\begin{figure}[!t]
\centerline{
\psfig{figure=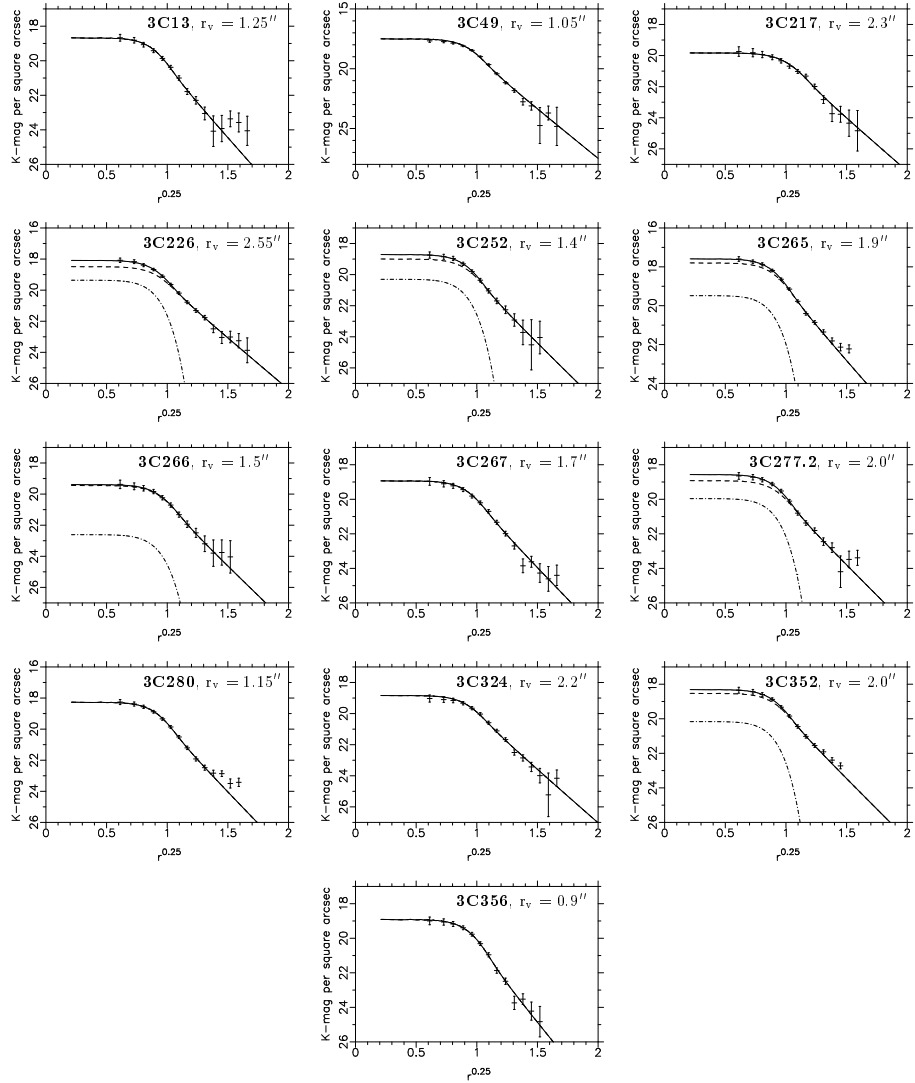,clip=,width=\textwidth}
}
\caption{\label{irvauc} Fits to the radial intensity profiles of the
K--band images of the 13 3CR galaxies for which enhanced optical emission
prevented such a fit to the optical intensity profile. The characteristic
radii have been determined from the best fitting profiles. The same
notation is used as in Figure~\ref{kvaucs}.}
\end{figure}

Such fits could also be made for the remaining galaxies in the sample, for
which no de Vaucouleurs fit to the HST data was possible due to the
aligned emission. For these galaxies, the characteristic radius was
allowed to be a further free parameter in the fit. For the five galaxies
with redshifts $z > 1.4$, the low signal--to--noise ratio of the infrared
images meant that the $\chi ^2$ of the fit varied only slowly with
characteristic radius and so the best--fitting characteristic radius was
not well--defined; these galaxies have therefore been omitted from the
analysis. For the remaining galaxies the characteristic radius could
typically be determined to an accuracy of $\sim 35\%$, and the best
fitting models are shown in Figure~\ref{irvauc}.

Figures~\ref{kvaucs} and~\ref{irvauc} demonstrate that, except for the two
cases of 3C22 and 3C41, there is only a small ($< 10\%$) point source
component contributing to the total K--band flux density of the 3CR radio
galaxies. Indeed, the point source contribution is consistent, within the
90\% confidence limits, with being zero in all but these two cases. For
3C22 and 3C41, approximately 37\% and 24\% (respectively) of the K--band
emission is associated with an unresolved emission source. It is
interesting to note that these two galaxies are the brightest in our
sample in the K--band, lying furthest from the mean K$-z$ relationship
(Best \etal\ 1997a).  Rawlings \etal\ (1995) have previously proposed that
3C22 possesses a significant unresolved K--band contribution and, together
the detection of broad \Ha\ emission (Economou \etal\ 1995) from this
galaxy, this suggests that this source may be a reddened quasar.

\smallskip

\begin{figure}[t]
\centerline{
\psfig{figure=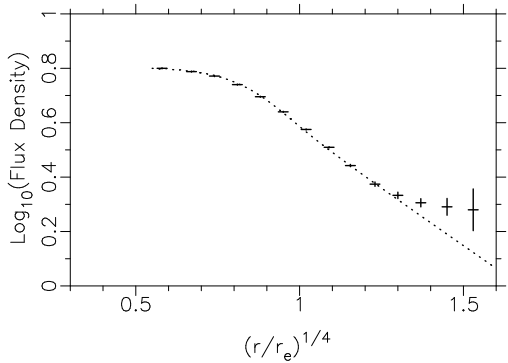,angle=-90,clip=,width=6.7cm}
}
\caption{\label{vaucsum} A combined radial intensity profile in the
K--band for the 12 galaxies with $1.0 \le r_{\rm e} \le 2.0$. Units on the
y--axis are arbitrary. Errors for each point are marked. The dotted line
shows a combined de Vaucouleurs profile: halo emission is clearly visible
at $(r/r_{\rm e})^{1/4} \gta 1.25$.}
\end{figure}

It is noticeable that for a proportion of the galaxies the observed
emission profile becomes brighter than the predicted de Vaucouleurs
profile at large radii, suggesting that these galaxies possess diffuse
extended envelopes.  To improve the signal--to--noise of this feature, we
have scaled the radial profile of each galaxy with respect to the
characteristic radius derived for it, and summed the intensity profiles.
Only those galaxies for which $1.0 \le r_{\rm e} \le 2.0$ were included:
for galaxies with a smaller $r_{\rm e}$ there will still be a significant
effect due to the seeing at radii of 2 to 3 $r_{\rm e}$, whilst for
galaxies with a larger $r_{\rm e}$ there is insufficient signal at the
largest radii to accurately test for the presence of the halo. For the 12
galaxies which meet this criteria, the scaled radial profiles were summed,
weighting each galaxy equally, and the results are presented in
Figure~\ref{vaucsum}. At radii $\left (r/r_{\rm e} \right )^{1/4} \gta
1.25$, corresponding to $r \gta 35$\,kpc, there is a clearly significant
halo component. 

This halo component is very similar to that of low redshift cD galaxies,
which lie towards the centre of galaxy clusters. Together with the large
(a few times $10^{11} M_{\odot}$) masses of these galaxies, this suggests
that high redshift 3CR galaxies may live in moderately rich
environments. Existing evidence supports this hypothesis, eg. the
detection of cooling flows around these galaxies (Crawford and Fabian
1995), evidence for companion galaxies in narrow--band [OII] images
(McCarthy 1988), and the fact that individual well--studied 3CR galaxies
at high redshift appear to live in clusters (eg. 3C324, Dickinson
\etal\ 1996).

\section{The Kormendy relation for the redshift one 3CR radio galaxies}

The Kormendy $r_{\rm e}$ $vs$ $\mu_{\rm e}$ projection of the fundamental
plane alleviates the need for detailed spectroscopy, which is a difficult
process for high redshift galaxies. In the previous section we obtained
good estimates for the characteristic radii of the 3CR galaxies; by
measuring also their surface brightnesses, it will be possible to compare
their location on the Kormendy projection with those of low redshift
brightest cluster galaxies and giant ellipticals. Thus, we can investigate
how much evolution of the stellar populations must be occurring with
cosmic epoch.

There are a number of details which need to be dealt with before the
location of the 3CR galaxies on the Kormendy relation can be
determined. The foremost of these is that we can not use the HST images to
measure the surface brightnesses, because of the large component of light
associated with the aligned emission at these wavelengths. Instead, the
K--band images must be used, with the adoption of an appropriate
k--correction. We must also take account of the effects of seeing in the
K--band observations, of any point source or flat spectrum contributions
to the flux density (as calculated in the earlier sections of this paper),
and compensate for cosmological surface brightness dimming.

\begin{figure}[!t]
\centerline{
\psfig{figure=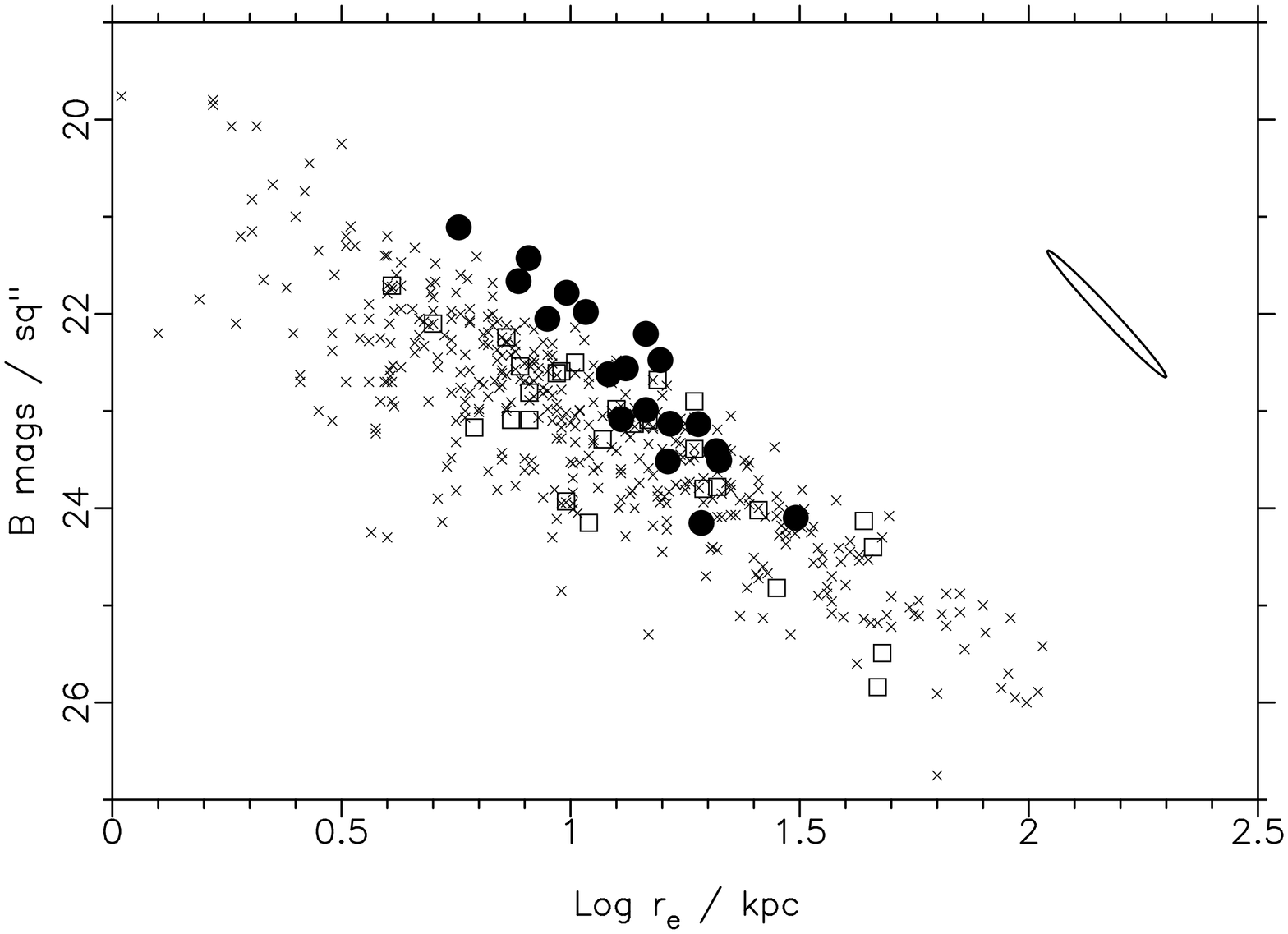,angle=-90,clip=,width=8.8cm}}
\centerline{
\psfig{figure=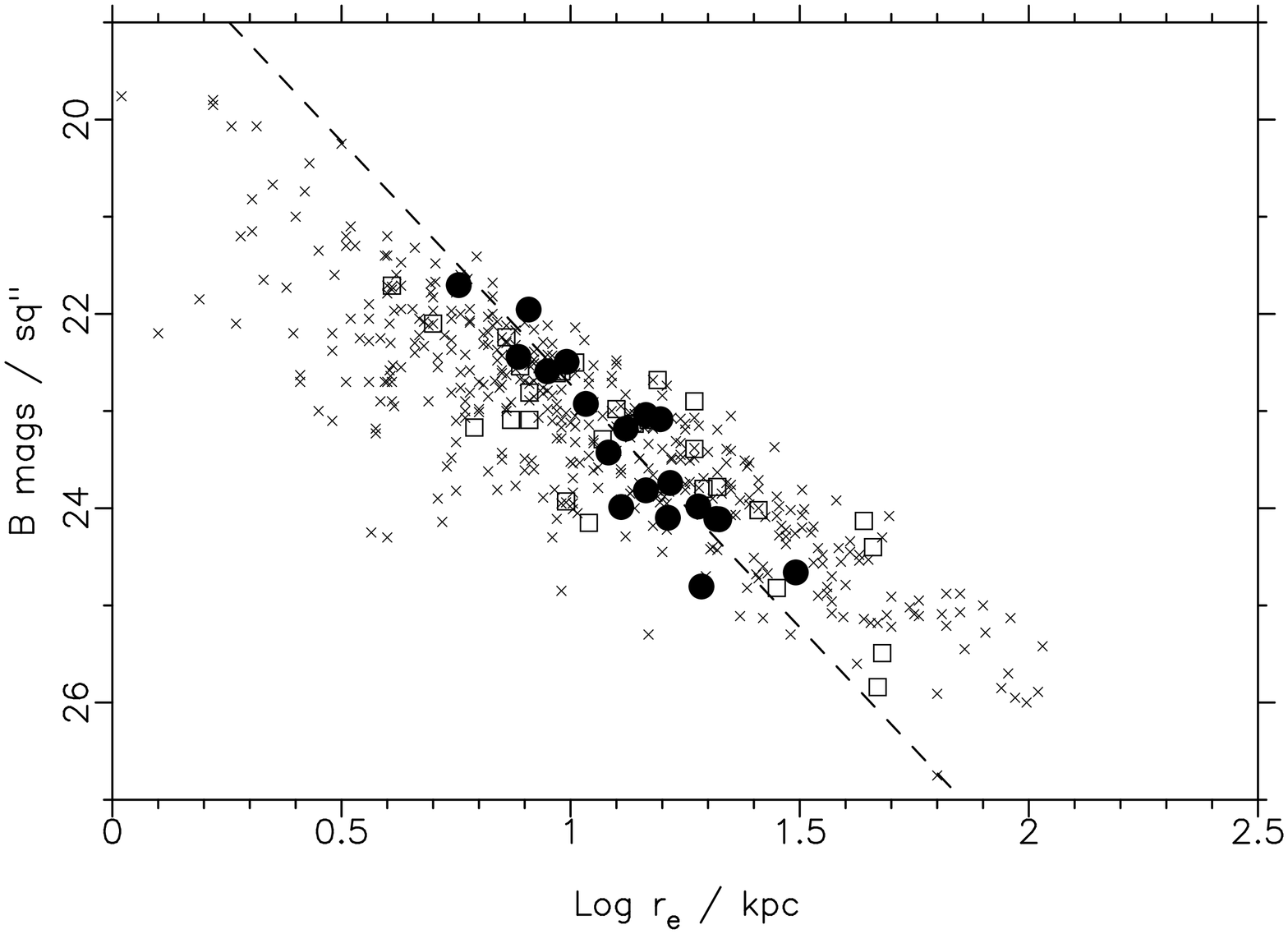,angle=-90,clip=,width=8.8cm}}
\caption{\label{revsmu} Plots of B--band surface brightness vs
characteristic radius for the 3CR galaxies (solid circles) compared with
low redshift giant ellipticals and brightest cluster galaxies (crosses),
and low redshift radio galaxies (open squares). The low redshift data are
taken from Rigler and Lilly (1994) and references therein. The ellipse
indicates the error ellipse for the 3CR galaxies whose characteristic
radius was measured from the UKIRT data; those for which the measurement
was from the HST data have much an error ellipse less than half of this
size. (a) Assuming no evolution of the stellar populations of the 3CR
galaxies. (b) Assuming that the stellar populations evolve passively. The
dashed line shows a line of constant total luminosity.}
\end{figure}

As a first, null hypothesis, we assume that the stellar populations of the
3CR radio galaxies are not evolving, and so we use the SED of low redshift
giant elliptical galaxies to calculate the required k--correction. The
rest--frame B--band surface brightnesses thus calculated for the 3CR radio
galaxies are plotted against the measured de Vaucouleurs radii in
Figure~\ref{revsmu}a, together with data from low redshift samples of
giant ellipticals and brightest cluster galaxies (Oegerle and Hoessel
1991, Schombert 1987) and a sample of low redshift radio galaxies (Lilly
and Prestage 1987). The 3CR galaxies have greater surface brightnesses
than the low redshift galaxies in the fundamental plane, implying that
some stellar evolution must have occurred between a redshift of one and
the current epoch.

We can repeat the process, but instead make the more physical assumption
that the stellar populations of the 3CR radio galaxies form at high
redshift ($z=10$) and then evolve passively. If evolution is occurring,
the stellar populations observed in the high redshift radio galaxies will
be younger, and hence brighter, than those seen in the nearby
galaxies. The Bruzual and Charlot (1993, 1997) stellar synthesis codes were
used to construct such passively evolving galaxies with the age that each
3CR galaxy would possess, and thus to calculate the required
k--correction.  The models were then used to determine the evolution in
the rest--frame B--magnitude of the stellar populations that would occur
for each galaxy between its observed redshift and a redshift of zero. This
procedure therefore derived surface brightnesses of the 3CR galaxies that
could be compared directly with low redshift giant ellipticals in the
fundamental plane.

The derived surface brightnesses are plotted against the de Vaucouleurs
radius in Figure~\ref{revsmu}b. It can be seen that they lie along the
fundamental plane defined by the low redshift giant ellipticals, providing
further evidence that the stellar populations evolve passively. There is
an indication that the slope of the fundamental plane defined by the 3CR
galaxies may be slightly steeper than that of the low redshift giant
ellipticals, but this is likely to be just a selection effect: the dashed
line on Figure~\ref{revsmu}b shows a line of constant total luminosity for
the galaxies, that is, a line along which the product of the surface
brightness and the square of the characteristic radius is constant; the
3CR galaxies lie closely along this line. That is, they lie within the
fundamental plane, and also along the line of constant luminosity.

According to standard cannibalism models (eg. Hausman and Ostriker 1978)
the position of a galaxy along the fundamental plane is interpreted as
being related to its merger history. Galaxies which have undergone more
mergers and are more highly evolved lie further to the right along the
fundamental plane. In this respect it is interesting to note that: (i) on
Figure~\ref{revsmu}, the high redshift 3CR galaxies possess a smaller
spread of characteristic radii than those of the low redshift samples,
indicating perhaps that these galaxies are all seen at a similar point in
their evolutionary history, and (ii) the mean characteristic radius of the
3CR galaxies ($14.6 \pm 1.4$\,kpc) is larger than that of the low redshift
samples ($11.0 \pm 0.5$\,kpc). The indicates that the 3CR galaxies must be
highly evolved systems, even by a redshift of one.

\section{Conclusions}

From our study of the high redshift 3CR radio galaxies, we conclude the
following points:

\begin{bul}
\item The radial intensity profiles of the host galaxies of the 3CR radio
sources are well matched by de Vaucouleurs' law.
\item The galaxies possess extended cD type halos. 
\item Passive evolution of the stellar populations is required if the 3CR
galaxies are to lie along the fundamental plane defined by low redshift giant
ellipticals. 
\item Their characteristic radii are large, with relatively little scatter.
\end{bul}

We conclude that the 3CR radio galaxies at redshift $z \sim 1$ are highly
evolved galaxies, containing old stellar populations, which lie at the
centre of moderately rich (proto--)clusters.


\end{document}